\documentclass[aps,preprint]{revtex4}%
\usepackage{amsfonts}
\usepackage{amsmath}
\usepackage{amssymb}
\usepackage{graphicx}%
\providecommand{\U}[1]{\protect\rule{.1in}{.1in}}

\begin{document}
\title[Hawking Radiation of Grumiller Black Hole ]{Hawking Radiation of Grumiller Black Hole }
\author{S.F. Mirekhtiary}
\email{fatemeh.mirekhtiary@emu.edu.tr}
\author{I. Sakalli}
\email{izzet.sakalli@emu.edu.tr}
\affiliation{Department of Physics, Eastern Mediterranean University, Gazimagosa, North
Cyprus, Mersin 10, Turkey }
\keywords{Hawking radiation, relativistic Hamilton--Jacobi equation, Grumiller black
hole, quantum tunneling, back reaction effect, pioneer anomaly. }
\pacs{PACS number}

\begin{abstract}
In this paper, we consider the relativistic Hamilton-Jacobi \ (HJ) equation
and study the Hawking radiation (HR) of scalar particles from uncharged
Grumiller black hole (GBH) which is affordable for testing in astrophysics.
GBH is also known as Rindler modified Schwarzschild BH. Our aim is not only to
investigate the effect of the Rindler parameter $a$ on the Hawking temperature
($T_{H}$), but to examine whether there is any discrepancy between the
computed horizon temperature and the standard $T_{H}$ as well. For this
purpose, in addition to its naive coordinate system, we study on the three
regular coordinate systems which are Painlev\'{e}-Gullstrand (PG), ingoing
Eddington-Finkelstein (IEF) and Kruskal-Szekeres (KS) coordinates. In all
coordinate systems, we calculate the tunneling probabilities of incoming and
outgoing scalar particles from the event horizon by using the HJ equation. It
has been shown in detail that the considered HJ method is concluded with the
conventional $T_{H}$ in all these coordinate systems without giving rise to
the famous factor-2 problem. Furthermore, in the PG coordinates
Parikh-Wilczek's tunneling (PWT) method is employed in order to show how one
can integrate the quantum gravity (QG) corrections to the semiclassical
tunneling rate by including the effects of self-gravitation and back reaction.
We then show how these corrections yield a modification in the $T_{H}$.

\end{abstract}
\volumeyear{2014}
\volumenumber{61}
\issuenumber{5}
\eid{Communications in Theoretical Physics}
\date[Date text]{date}
\received[Received text]{date}

\revised[Revised text]{date}

\accepted[Accepted text]{date}

\published[Published text]{1 May 2014 }

\startpage{1}
\endpage{ }
\maketitle
\tableofcontents

\section{INTRODUCTION}

Rindler acceleration \cite{Rindler}, which acts on an observer accelerated in
a flat spacetime has recently become rage anew. This is due to its similarity
with the mysterious acceleration that revealed after the long period
observations on the Pioneer spacecraft -- Pioneer 10 and Pioneer 11 -- after
they covered a distance about $3\times10^{9}km$ on their paths out of the
Solar System \cite{Berman}. Contrary to the expectations, that mysterious
acceleration is attractive i.e., directed toward the Sun and this phenomenon
is known as the Pioneer anomaly. Firstly, Grumiller \cite{Grumiller0} (and
later together with his collaborators \cite{Grumiller12,Grumiller3}) showed
the correlation between the $a$ and the Pioneer anomaly. On the other hand,
Turyshev et al. \cite{Turyshev} have recently made an alternative study to the
Grumiller's ones in which the Pioneer anomaly is explained by thermal heat
loss of the satellites.

Another intriguing feature of the $a$ is that it may play the role of dark
matter in galaxies \cite{Grumiller0,Grumiller12}. Namely, the incorporation of
the Newton's theory with the $a$ might serve to explain rotation curves of
spiral galaxies without the presence of a dark matter halo (a reader may refer
to the study of Lin et al. \cite{NanLin}). For the galaxy-Sun pair, the $a$
with the order $\sim10^{-11}m/s^{2}$ in physical units is a very close value
to MOND's acceleration which successfully describes rotation curves without a
dark matter halo (see \cite{Mastache} and references therein). However, very
recently the studies \cite{NanLin,Mastache} have been retested and criticized
by Cervantes-Cota and G\'{o}mez-L\'{o}pez \cite{Cota}.

As stated in \cite{Grumiller0,Grumiller12}, the main function of the $a$ is to
constitute a crude model which casts doubts on the description of rotation
curves with a linear growing of the velocity with the radius. By virtue of
this, in the novel study of \cite{Grumiller0} it was suggested that the
effective potential of a point mass $M$ should include $r-$dependent
acceleration term. Moreover, in the studies of \cite{Grumiller0,Grumiller12}
it is explicitly shown that dilatonic field source in general relativity (GR)
is required for deriving a spacetime with the $a$. However, in performing this
process one should be cautious about the physical energy conditions. It has
been recently revived by \cite{Halilsoy1} that the GBH spacetime
\cite{Grumiller12} does not satisfy the all essential energy conditions of the
GR. Very recently, Mazharimousavi and Halilsoy (MH) \cite{Halilsoy2}\ have
shown that the GBH metric becomes physically acceptable in the $f(\Re)$
gravity. In other words, in the $f(\Re)$ gravity the problematic energy
conditions are all fixed. The physical source that has been used in
\cite{Halilsoy2} possesses a perfect fluid-type energy momentum tensor, and
the pressure of the fluid becomes negative with a particular choice. So, one
can infer that the $a$ plays the role of the dark matter. From now on, we
designate the metric of \cite{Halilsoy2}\ \ as
Grumiller-Mazharimousavi-Halilsoy BH and abbreviate it as GMHBH. Meanwhile,
very recently detailed analysis of the geodesics of this BH has been made by
\cite{HOM}.

More than forty years ago, Bekenstein unraveled that the entropy of a BH is
proportional to its surface area {\cite{Bekenstein1,Bekenstein2,Bekenstein3}}.
Afterwards Hawking studied the particle creation around the event horizon of a
BH to ascertain that the BH possesses a black body type thermal radiation with
the temperature subject to its surface gravity \cite{Hawking1,Hawking2}. After
these novel studies of Hawking, up to date there is a rapidly growing
literature on the thermodynamics of various kinds of BHs. Moreover, deriving
alternative methods to the HR which divulges\ the underlying BH\ spacetime has
always remained on the agenda. For the topical review of the HR together with
its available methods, a reader may consult to \cite{DonPage}. Among those
alternative methods for the HR, utilization of the relativistic HJ equation is
one of the runproof methods. This method is developed by \cite{Angheben} that
basically employs the complex path analysis of Padmanabhan et al.
\cite{Padnab1,Padnab2,Padnab3}. The associated method involves the WKB
approximation and calculates the imaginary part of the action of the tunneling
particles. In performing this process one should ignore the self-gravitational
effects of the tunneling particle and the energy conservation. In general, the
relativistic HJ equation can be solved by substituting a suitable ansatz. For
the separability of the equation the chosen ansatz should take account of the
Killing vectors of the spacetime. Thus we obtain an integral equation which
yields the classically forbidden trajectory that starts from inside of the BH
and finishes at the outside observer. On the other hand, the integral under
question has always a pole located at the event horizon of the BH. We recall
that such integrals are evaluated by applying the method of complex path
analysis in order to circumvent the pole. Result of the integral leads us to
get the tunneling rate for the GMHBH which renders possible to read the
$T_{H}$. On the other hand, PWT method \cite{KW1,KW2,PW} uses the null
geodesics to derive the $T_{H}$ as a quantum tunneling process. In this
method, self-gravitational interaction of the radiation and energy
conservation are taken into account. As a result, the HR spectrum can not be
strictly thermal for many well-known BHs, like Schwarzschild,
Reissner-Nordstr\"{o}m etc. \cite{PW,Zhang}.

Here we plan to investigate the HR of a static and spherically symmetric GMHBH
via the well-known HJ and PWT methods. We restate that, in this paper, we
shall make only an application of the associated methods to the GMHBH. By
doing this, we aim not only to make an analysis about the influences of the
$a$ on the HR, but to test whether the associated methods employing for the
GMHBH with different coordinates yield the true $T_{H}$\ without admitting the
factor-2 problem or not. For the review of the factor-2 problem arising in the
HR, a reader may refer to
\cite{Akhmedov,Akhmedov2,Pilling,BZhang,Vanzo,GRG12,SM13}.

First of all, we shall review the GMHBH which has a fluid source in the
context of $f(\Re)$ gravity \cite{Halilsoy2}. Then we use the HJ method in
order to calculate the imaginary part of the classical action for outgoing
trajectories crossing the horizon. In addition to the naive coordinates, three
more coordinate systems (all regular) which are PG, IEF and KS, respectively,
are considered. Slightly different from the other coordinate systems, during
the application of the HJ method in the KS coordinates, we will first reduce
the GMHBH spacetime to a Minkowski type space with a conformal factor, and
then show in detail how one recovers the $T_{H}$. Furthermore, in the PG
coordinate system we shall study the PWT method in order to give a QG
correction to the tunneling probability by considering the back reaction
effect. To this end, the log-area correction to the Bekenstein-Hawking entropy
will be taken into account. Finally, the modified $T_{H}$ due to the back
reaction effect will be computed.

The paper uses the signature $(-,+,+,+)$ and the geometrical units
$c=G=\hbar=k_{B}=1$. The paper is organized as follows. In Sec. II, we review
some of the geometrical and thermodynamical features of the GMHBH. We also
show how the HJ equation is separated by a suitable ansatz within the naive
coordinates. The calculations of the tunneling rate and henceforth the $T_{H}$
via the HJ method are also represented. In Sec. III the HR of the GMHBH in the
PG coordinates is analyzed in the frameworks of the HJ and PWT methods. The
back reaction effect on the $T_{H}$ is also examined. Sec. IV and V are
devoted to the application of the HJ method in the IEF and KS coordinate
systems, respectively. Finally, the conclusion and future directions are given
in Sec. VI.

\section{GMHBH AND HJ METHOD}

In this section we will first present the geometry and some thermodynamical
properties of the GMHBH. Then, with aid of a suitable ansatz we will get the
radial equation for the relativistic HJ equation in the background of the
GMHBH. Finally, we represent how the HJ method culminates in the $T_{H}$.

The $4D$ action obtained from $f(\Re)$ gravity is given by%

\begin{equation}
S=\frac{1}{2\lambda}\int\sqrt{-g}f(\Re)d^{4}x+S_{M}, \label{1}%
\end{equation}

where $\lambda=8\pi G=1,$ $\Re$ is the curvature scalar and $f(\Re)=\Re
-12a\xi\ln\left\vert \Re\right\vert $ in which $a$ and $\xi$ are positive
constants. $S_{M}$ denotes the physical source for a perfect fluid-type energy
momentum tensor%

\begin{equation}
T_{\mu}^{\upsilon}=diag.[-\rho,p,q,q], \label{2}%
\end{equation}

with the thermodynamic pressure $p$ being a function of the rest mass density
of the matter (for short: matter density) $\rho$ only, so that $p=-\rho$.
Meanwhile, $q$ is also a state function which is to be determined. Recently,
MH has obtained the GMHBH solution to the above action in their landmark paper
\cite{Halilsoy2}. Their solution is described by the following $4D$ static and
spherically symmetric line element%
\begin{equation}
ds^{2}=-Hdt^{2}+\frac{dr^{2}}{H}+r^{2}d\Omega^{2}, \label{3}%
\end{equation}

where $d\Omega^{2}$ is the standard metric on $2-$sphere and the metric
function $H(r)$ is computed as
\begin{equation}
H=1-\frac{2M}{r}+2ar=\frac{2a}{r}(r-r_{h})(r-r_{0}), \label{4nn}%
\end{equation}

which is nothing but the metric function of the GBH without the cosmological
constant \cite{Grumiller0}. Here, $M$ represents the constant mass and
\begin{equation}
r_{0}=-\frac{\sqrt{1+16aM}+1}{4a}, \label{5nn}%
\end{equation}

which cannot be horizon due to its negative signature. Therefore, the GMHBH
possesses only one horizon (event horizon, $r_{h}$) which is given by%

\begin{equation}
r_{h}=\frac{\sqrt{1+16aM}-1}{4a}, \label{6nn}%
\end{equation}

Further, it is found that the energy-momentum components are%

\begin{equation}
p=-\rho=\frac{\left[  6a\xi-f(\Re)\right]  r^{2}+4(\xi-a)r-6M\xi}{2r^{2}},
\label{7nn}%
\end{equation}

\begin{equation}
q=-\frac{f(\Re)r-2\xi+8a}{2r}, \label{8nn}%
\end{equation}

where%

\begin{equation}
f(\Re)=-\left[  \frac{12a}{r}+12a\xi\ln\left(  \frac{12a}{r}\right)  \right]
, \label{9nn}%
\end{equation}

One can easily observe from the last three equations that the $a$ is decisive
for the fluid source. This can be best seen by simply taking the limit of
$a\rightarrow0$ which corresponds to the vanishing fluid and Ricci scalar, and
so forth $\xi\rightarrow0.$ In short, $f(\Re)$ gravity reduces to the usual
$\Re$-gravity. In short, while $a\rightarrow0$ the GMHBH reduces to the
well-known Schwarzschild BH.

Surface gravity \cite{Wald} of the GMHBH can simply be calculated through the
following expression%

\begin{equation}
\kappa(M)=\left.  \frac{H^{\prime}}{2}\right\vert _{r=rh}=\frac{a\left(
r_{h}-r_{0}\right)  }{r_{h}}, \label{10nn}%
\end{equation}

where a prime "$\prime$" denotes differentiation with respect to $r$. From
here on in, one obtains the Hawking temperature of the GMHBH as%

\begin{align}
T_{H}  &  =\frac{\kappa(M)}{2\pi}=\frac{a\left(  r_{h}-r_{0}\right)  }{2\pi
r_{h}},\nonumber\\
&  =\frac{a\sqrt{1+16aM}}{\pi\left(  \sqrt{1+16aM}-1\right)  }, \label{11}%
\end{align}

From the above expression, it is seen that while the GMHBH losing its $M$ by
virtue of the HR, $T_{H}$ increases (i.e., $T_{H}\rightarrow\infty$) with
$M\rightarrow0$ in such a way that its divergence speed is tuned by $a$.
Meanwhile, one can check that $\lim_{a\rightarrow0}T_{H}=\frac{1}{8\pi M}$
which is well-known Hawking temperature computed for the Schwarzschild BH. The
Bekenstein-Hawking entropy is given by%

\begin{equation}
S_{BH}=\frac{A_{h}}{4}=\pi r_{h}^{2}, \label{12nn}%
\end{equation}

Its differential form is written as%

\begin{equation}
dS_{BH}=\frac{4\pi}{\sqrt{1+16aM}}r_{h}dM, \label{13nn}%
\end{equation}

By using the above equation, the validity of the first law of thermodynamics
for the GMHBH can be approved via%

\begin{equation}
T_{H}dS_{BH}=dM. \label{14nn}%
\end{equation}

Here, we consider the problem of a scalar particle (spin-0) which crosses the
event horizon from inside to outside while there is no back-reaction effect
and self-gravitational interaction. Within the semi-classical framework, the
classical action $I$ of the particle satisfies the relativistic HJ equation
\cite{Angheben}\ is given by%

\begin{equation}
g^{\mu\nu}\partial_{\mu}I\partial_{\nu}I+m^{2}=0, \label{15nn}%
\end{equation}

in which $m$ is the mass of the scalar particle, and $g^{\mu\nu}$ represents
the invert metric tensors derived from the metric (3). By considering Eqs.
(3), (4) and (15), we get%

\begin{equation}
\frac{-1}{H}(\partial_{t}I)^{2}+H(\partial_{r}I)^{2}+\frac{1}{r^{2}}%
(\partial_{\theta}I)^{2}+\frac{1}{r^{2}\sin^{2}\theta}(\partial_{\varphi
}I)^{2}+m^{2}=0, \label{16nn}%
\end{equation}

For the HJ equation it is general to use the separation of variables method
for the action $I=I(t,r,\theta,\varphi)$ as follows%

\begin{equation}
I=-Et+W(r)+J(x^{i}), \label{17nn}%
\end{equation}

where%

\begin{equation}
\partial_{t}I=-E,\text{ \ \ }\partial_{r}I=\partial_{r}W(r),\text{
\ \ }\partial_{i}I=J_{i}, \label{18n}%
\end{equation}
and $J_{i}$'s are constants in which $i=1,2$ identifies angular coordinates
$\theta$ and $\varphi$, respectively. The norm of the timelike Killing vector
$\partial_{t}$ becomes (negative) unity at a particular location:%

\begin{equation}
r\equiv R_{d}=\frac{r_{h}+r_{0}}{2}+\frac{1+\sqrt{4(r_{h}-r_{0})^{2}%
+4(r_{h}+r_{0})a+1}}{4a}, \label{19n}%
\end{equation}

It means that when a detector of an observer is located at $R_{d}$ which is
outside the horizon, the energy of the particle measured by the observer is
$E$. Solving Eq. (16) for $W(r)$ yields%

\begin{equation}
W(r)=\pm\int\frac{\sqrt{E^{2}-\frac{H}{r^{2}}\left(  J_{\theta}^{2}%
+\frac{J_{\varphi}^{2}}{\sin^{2}\theta}+m^{2}r^{2}\right)  }}{H}dr,
\label{20n}%
\end{equation}

The quadratic form of Eq. (16) is the reason of $\pm$ signatures that popped
up in the above equation. Solution of Eq. (20) with "$+$" signature
corresponds to the outgoing scalar particles and the other solution i.e., the
solution with "$-$" signature refers to the ingoing particles. Evaluating the
above integral around the pole at the horizon (following to the prescription
given by \cite{Feynman}), one reaches to%

\begin{equation}
W_{\left(  \pm\right)  }=\pm\frac{i\pi Er_{h}}{2a(r_{h}-r_{0})}+\delta,
\label{21n}%
\end{equation}

where $\delta$ is a complex integration constant. Thus, we can deduce that
imaginary parts of the action arises due to the pole at the horizon and from
the complex constant $\delta$. Thence, we can determine the probabilities of
ingoing and outgoing particles while crossing $r_{h}$ as%

\begin{equation}
P_{out}=e^{-2\operatorname{Im}I}=\exp\left[  -2{\operatorname{Im}}W_{\left(
+\right)  }\right]  , \label{22n}%
\end{equation}

\begin{equation}
P_{in}=e^{-2\operatorname{Im}I}=\exp\left[  -2{\operatorname{Im}}W_{\left(
-\right)  }\right]  , \label{23n}%
\end{equation}

In the classical point of view, a BH absorbs any ingoing particles passing its
horizon. In other words, there is no reflection for the ingoing waves which
corresponds to $P_{in}=1$. This is enabled by setting ${\operatorname{Im}%
}\delta=\frac{\pi Er_{h}}{2a(r_{h}-r_{0})}.$ This choice also implies that the
imaginary part of the action $I$ for a tunneling particle can only come out
$W_{(+)}$. Namely, we get
\begin{equation}
\operatorname{Im}I={\operatorname{Im}}W_{(+)}=\frac{\pi r_{h}E}{a(r_{h}%
-r_{0})}. \label{24n}%
\end{equation}

Therefore, the tunneling rate for the GMHBH can be obtained as%

\begin{equation}
\Gamma=P_{out}=e^{\frac{-2\pi Er_{h}}{a(r_{h}-r_{0})}}, \label{25n}%
\end{equation}

and according to \cite{PW}%

\begin{equation}
\Gamma=e^{-\beta E}, \label{26n}%
\end{equation}

in which $\beta$ denotes the Boltzmann factor and $T=\frac{1}{\beta}$, one can
easily read the horizon temperature of the GMHBH as
\begin{equation}
\check{T}_{H}=\frac{a(r_{h}-r_{0})}{2\pi r_{h}}. \label{27n}%
\end{equation}

This nothing but the $T_{H}$ obtained in Eq. (11).

\section{HJ AND PWT METHODS WITHIN PG COORDINATES}

In the literature, PG coordinates are known as the first coordinate system
which is non-singular at the event horizon and allow us to describe timelike
or null worldlines inward crossing the horizon. In other words, we use the PG
coordinates \cite{Painleve,Gullstrand} in order to describe the spacetime on
either side of the event horizon of a static BH. In this coordinate system,
the generic spherically metric (3) loses its diagonal or static form. Instead
it allows a cross term which makes the metric stationary and no longer
symmetric, but oriented. Thus, an observer does not consider the surface of
the horizon to be in any way special. In this section, we consider the PG
coordinates of the GMHBH not only in the HJ method, but in the PWT method as
well. Then we show how both methods yield the $T_{H}$. Besides, the back
reaction effect on the $T_{H}$ is thoroughly discussed.

We can pass to the PG coordinates by applying the following transformation
\cite{Robertson} to the metric (3)%

\begin{equation}
dt_{PG}=dt+\frac{\sqrt{1-H}}{H}dr, \label{28n}%
\end{equation}

where $t_{PG}$ is our new time coordinate (let us call it as PG time). One of
the main properties of these coordinates is that $t_{PG}$ concurrently
corresponds to the proper time. After substituting Eq. (28) into the metric
(3), one obtains the PG line-element as follows
\begin{equation}
ds^{2}=-Hdt_{PG}^{2}+2\sqrt{1-H}dt_{PG}dr+dr^{2}+r^{2}d\Omega^{2}, \label{29n}%
\end{equation}

For the metric (29), the HJ equation (15) becomes%

\begin{equation}
-(\partial_{t_{PG}}I)^{2}+2\sqrt{1-H}(\partial_{t_{PG}}I)(\partial
_{r}I)+H(\partial_{r}I)^{2}+\frac{1}{r^{2}}(\partial_{\theta}I)^{2}+\frac
{1}{r^{2}\sin^{2}\theta}(\partial_{\varphi}I)^{2}=0 \label{30nn}%
\end{equation}

Letting%

\begin{equation}
I=-Et_{PG}+W_{PG}(r)+J(x^{i}), \label{31n}%
\end{equation}

and now by substituting for the above ansatz in Eq. (30), we obtain%

\begin{equation}
W_{PG}(r)=\int\frac{E\sqrt{1-H}}{H}\left(  1\pm\sqrt{1-\frac{HF }{\left(
1-H\right)  E^{2}}}\right)  dr, \label{32n}%
\end{equation}
where%

\begin{equation}
F=m^{2}-E^{2}+\frac{J_{\theta}^{2}}{r^{2}}+\frac{J_{\varphi}^{2}}{r^{2}%
\sin^{2}\theta}, \label{33n}%
\end{equation}

Thus one can see that near the horizon Eq. (32) reduces to%

\begin{equation}
W_{PG(\pm)}=E\int\frac{1}{H}(1\pm1)dr, \label{34n}%
\end{equation}

Since $W_{PG(-)}=0$ which is a warranty condition for non-reflection of the
ingoing particles, we thus have%

\begin{equation}
W_{PG(+)}=\frac{ir_{h}\pi E}{a(r_{h}-r_{0})}. \label{35n}%
\end{equation}

So, we get the imaginary part of the $I$ as%

\begin{equation}
\operatorname{Im}I={\operatorname{Im}}W_{PG(+)}=\frac{\pi r_{h}E}%
{a(r_{h}-r_{0})}, \label{36n}%
\end{equation}

After recalling Eqs. (25) and (26), we can readily read the horizon
temperature of the GMHBH which is expressed in the PG\ coordinates as%

\begin{equation}
\check{T}_{H}=\frac{a(r_{h}-r_{0})}{2\pi r_{h}}. \label{37n}%
\end{equation}

This result is full measure of the standard value of the $T_{H}$ (11).

Now, employing the tunneling method prescribed by \cite{PW} we recalculate the
imaginary part of the $I$ for an outgoing positive energy particle which
crosses the horizon outwards in the PG coordinates. In the metric (29),\ the
radial null geodesics of a test particle has a rather simple form%

\begin{equation}
\dot{r}=\frac{dr}{d_{t_{PG}}}=-\sqrt{1-H}\pm1, \label{38}%
\end{equation}

where upper (lower) sign corresponds to outgoing (ingoing) geodesics. After
expanding the metric function $H$ around the horizon $r_{h}$, we get%

\begin{equation}
H=H^{\prime}(r_{h})(r-r_{h})+O(r-r_{h})^{2}, \label{39}%
\end{equation}

and hence by using Eq. (10), the radial outgoing null geodesics, $\dot{r}%
$,\ can be approximately expressed as%

\begin{equation}
\dot{r}\cong\kappa(M)(r-r_{h}), \label{40}%
\end{equation}

The imaginary part of the $I$ for an outgoing positive energy particle which
crosses the horizon from inside ($r_{in}$) to outside ($r_{out}$) is given by%

\begin{equation}
\operatorname{Im}I=\operatorname{Im}\int_{r_{in}}^{r_{out}}p_{r}%
dr=\operatorname{Im}\int_{r_{in}}^{r_{out}}\int_{0}^{p_{r}}d\tilde{p}_{r}dr,
\label{41}%
\end{equation}

Hamilton's equation for the classical trajectory is given by%

\begin{equation}
dp_{r}=\frac{d\Pi}{\dot{r}}, \label{42}%
\end{equation}

where $p_{r}$ and $\Pi$\ denote radial canonical momentum and Hamiltonian,
respectively. So, one obtains%

\begin{equation}
\operatorname{Im}I=\operatorname{Im}\int_{r_{in}}^{r_{out}}\int_{0}^{\Pi}%
\frac{d\widetilde{\Pi}}{\dot{r}}dr, \label{43}%
\end{equation}

Now, if we consider the whole system as a spherically symmetric system of
total mass $M$, which is kept fixed, then this system consists of a GMHBH with
varying mass $M-\omega,$ emitting a spherical shell of mass $\omega$ such that
$\omega\ll M$. This phenomenon is known as self-gravitational effect
\cite{KW1}. After taking this effect into account, the above integration is
expressed as%

\begin{align}
\operatorname{Im}I  &  =\operatorname{Im}\int_{r_{in}}^{r_{out}}\int%
_{M}^{M-\omega}\frac{d\widetilde{\Pi}}{\dot{r}}dr,\nonumber\\
&  =-\operatorname{Im}\int_{r_{in}}^{r_{out}}\int_{0}^{\omega}\frac
{d\widetilde{\omega}}{\dot{r}}dr, \label{44}%
\end{align}

in which the Hamiltonian $\Pi=M-\omega$\ \textit{i.e.} $d\Pi=-d\omega$ is
used. Hence, $\dot{r}$ (40) can be reexperienced as follows%

\begin{equation}
\dot{r}\cong\kappa_{QG}(r-r_{h}), \label{45}%
\end{equation}

where $\kappa_{QG}=\kappa(M-\omega)$\ is the modified horizon gravity, which
is the so-called quantum gravity corrected surface gravity
\cite{ZhangKQG,Banerjee}. Thus, after $r$ integration (the integration over
$r$ is done by deforming the contour), Eq. (44) becomes%

\begin{equation}
\operatorname{Im}I=-\pi\int_{0}^{\omega}\frac{d\tilde{\omega}}{\kappa_{QG}},
\label{46}%
\end{equation}

So, let us express the "modified Hawking temperature" in the form of
$T_{QG}=\frac{\kappa_{QG}}{2\pi}$. From here on ,we get%

\begin{align}
\operatorname{Im}I  &  =-\frac{1}{2}\int_{0}^{\omega}\frac{d\tilde{\omega}%
}{T_{QG}},\nonumber\\
&  =-\frac{1}{2}\int_{S_{QG}(M)}^{S_{QG}(M-\omega)}dS,\nonumber\\
&  =-\frac{1}{2}\Delta S_{QG}, \label{47}%
\end{align}

then the modified tunneling rate is computed via%

\begin{equation}
\Gamma_{QG}\sim e^{-2\operatorname{Im}I}=e^{\Delta S_{QG}}. \label{48}%
\end{equation}

In string theory and loop quantum gravity, it is introduced with a logarithmic
correction (see for instance \cite{SQG,SQG2} and references therein)%

\begin{equation}
S_{QG}=\frac{A_{h}}{4}+\alpha\ln A_{h}+O(\frac{1}{A_{h}}), \label{49}%
\end{equation}

where $\alpha$ is a dimensionless constant, and it arises due to the back
reaction effects. It takes different values according to which theory is
considered \cite{SQG}. Thus, with the aid of Eqs. (12) and (49) one can
compute $\Delta S_{QG}$ as follows%

\begin{align}
\Delta S_{QG}  &  =-\frac{\pi\left(  8a\omega+\sqrt{1+16a(M-\omega)}%
-\sqrt{1+16aM}\right)  }{8a^{2}}+\nonumber\\
&  \alpha ln\left(  \frac{1+8a(M-\omega)-\sqrt{1+16a(M-\omega)}}%
{1+8aM-\sqrt{1+16aM}}\right)  , \label{50}%
\end{align}

Now, using the second law of thermodynamics%

\begin{equation}
T_{QG}dS_{QG}=dM, \label{51}%
\end{equation}

one can find the QG corrected form of the Hawking temperature $T_{QG}$\ due to
the back reaction. After a straightforward calculation, we can derive $T_{QG}$
from Eq. (51) in terms of the Hawking temperature as follows%

\begin{equation}
T_{QG}=\left(  1+\frac{\alpha}{\pi r_{h}^{2}}\right)  ^{-1}T_{H} \label{52}%
\end{equation}

Thus, one can easily see that once we ignore the back reaction effect (i.e.,
$\alpha=0$)\ we just produce the semiclassical Hawking temperature, $T_{H}$.
Meanwhile, it is also possible to obtain $T_{QG}$ from Eq. (48). For this
purpose, we expand $\Delta S_{QG}$ (50) and recast terms up to leading order
in $\omega$. So, one finds%

\begin{align}
\Delta S_{QG}  &  \cong-\left[  \frac{\pi}{a}\left(  \frac{\sqrt{1+16aM}%
)-1}{\sqrt{1+16aM}}\right)  +\frac{16a\alpha}{(1+16aM-\sqrt{1+16aM})}\right]
\omega+O(\omega^{2}),\nonumber\\
&  =-\left(  \frac{1}{T_{H}}+\alpha\frac{16\pi T_{H}}{1+16aM}\right)
\omega+O(\omega^{2}), \label{53}%
\end{align}

Based on Eqs. (26) and (48), we obtain%

\begin{equation}
\Gamma_{QG}\sim e^{\Delta S_{QG}}=e^{-\frac{\omega}{T}}, \label{54}%
\end{equation}

The inverse temperature, identified with the coefficient of $\omega$ is equal to%

\begin{equation}
T=\left(  \frac{1}{T_{H}}+\alpha\frac{16\pi T_{H}}{1+16aM}\right)  ^{-1}.
\label{55}%
\end{equation}

After manipulating the above equation, one can find that $T$ is nothing but
the QG corrected Hawking temperature (52). Namely, $T=T_{QG}.$

\section{HJ METHOD WITHIN IEF COORDINATES}

IEF coordinates are another regular coordinate system at the event horizon
which was originally constructed by \cite{Eddington,Finkelstein}. These
coordinates are aligned with radially moving photons. The generic metric (3)
takes the following form in the IEF coordinates (e.g. \cite{Poisson})%

\begin{equation}
ds^{2}=-Hd\upsilon^{2}+2\sqrt{1-H}d\upsilon dr+dr^{2}+r^{2}d\Omega^{2},
\label{56}%
\end{equation}

in which $\upsilon$ is a null coordinate which is the so-called advanced time.
It is given by%

\begin{equation}
\upsilon=t+r_{\ast}, \label{57}%
\end{equation}

where $r_{\ast}$ is known as the tortoise coordinate. For the outer region of
the GMHBH, it is found to be%

\begin{equation}
r_{\ast}=\int\frac{dr}{H}=\frac{1}{2a(r_{h}-r_{0})}\ln\left[  \frac{(\frac
{r}{r_{h}}-1)^{rh}}{(r-r_{_{0}})^{r_{0}}}\right]  , \label{58}%
\end{equation}

Since the metric (56) has a Killing vector field of $\xi^{\mu}=\partial
_{\upsilon}$, in this coordinate system an observer measures the scalar
particle's energy by $E=-\partial_{\upsilon}I$. In this regard, the action is
assumed to be of the form%

\begin{equation}
I=-E\upsilon+W_{EF}(r)+J(x^{i}). \label{59}%
\end{equation}

By using the above ansatz in the Eq. (15) for the metric (56), the final
expression for $W_{EF}(r)$ is found as%

\begin{equation}
W_{EF}(r)=\int\frac{E}{H}\left(  1\pm\sqrt{1-\frac{\tau H}{E^{2}}}\right)  dr,
\label{60}%
\end{equation}

in which%

\begin{equation}
\tau=m^{2}+\frac{J_{\theta}^{2}}{r^{2}}+\frac{J_{\varphi}^{2}}{r^{2}\sin
^{2}\theta}, \label{61}%
\end{equation}

Around the event horizon, we see that $W_{EF}(r)$ reduces to the following expression%

\begin{equation}
W_{EF(\pm)}=E\int\frac{1}{H}(1\pm1)dr, \label{62}%
\end{equation}

which is nothing but the same expression obtained in Eq. (34). Hereupon,
applying our standard procedure we get%

\begin{equation}
W_{EF(-)}=0,\enspace W_{EF(+)}=\frac{ir_{h}\pi E}{a(r_{h}-r_{0})}%
\rightarrow\,{\operatorname{Im}}I={\operatorname{Im}}W_{EF(+)}=\frac{\pi
Er_{h}}{a(r_{h}-r_{0})}, \label{63}%
\end{equation}

and likewise to Sec. IV, the horizon temperature computed for the GMHBH in the
IEF coordinates is of course that of the Hawking temperature:%

\begin{equation}
\check{T}_{H}=\frac{a(r_{h}-r_{0})}{2\pi r_{h}}=T_{H}. \label{64}%
\end{equation}

\section{HJ METHOD WITHIN KS COORDINATES}

Another non-singular coordinate system which covers the whole spacetime
manifold of the maximally extended BH solution is known as the KS coordinates
\cite{Kruskal,Szekeres}. These coordinates are generally used to properly
chart the spacetimes with the form of metric (3). Namely, the KS coordinates
are able to squeeze infinity into a finite distance, and thus the entire
spacetime can be visualized on a stamp-like diagram. In this section, we shall
employ the HJ equation for the KS form of the GMHBH in order to represent how
one gets the $T_{H}$ via the HJ method.

We can rewrite the metric (3) in the following form, as made in \cite{FRW},%

\begin{equation}
ds^{2}=-Hdudv+r^{2}d\Omega^{2}, \label{65}%
\end{equation}

where%

\begin{equation}
du=dt-dr_{\ast}\enspace dv=dt+dr_{\ast}, \label{66}%
\end{equation}

After defining new coordinates $(U,V)$ in terms of the surface gravity\ (10)
which are given by%

\begin{equation}
U=-e^{-\kappa u},\enspace V=e^{\kappa v}, \label{67}%
\end{equation}

we transform metric (65) to the KS form%

\begin{equation}
ds^{2}=\frac{H}{\kappa^{2}}\frac{dUdV}{UV}+r^{2}d\Omega^{2}. \label{68}%
\end{equation}

More explicitly, Eq. (68) becomes%

\begin{equation}
ds^{2}=-\pounds dUdV+r^{2}d\Omega^{2}, \label{69}%
\end{equation}

where%

\begin{equation}
\pounds =\frac{2r_{h}^{3}}{ar(r_{h}-r_{0})^{2}}(r-r_{0})^{1+\frac{r_{0}}%
{r_{h}}}, \label{70}%
\end{equation}

This metric is regular everywhere except at the physical singularity $r=0$.
Alternatively, the metric (69) can be transformed into%

\begin{equation}
ds^{2}=-\pounds (d\Im^{2}-dR^{2})+r^{2}d\Omega^{2}, \label{71}%
\end{equation}

which can be made by the following transformations%

\begin{equation}
\Im=\frac{1}{2}(V+U)=\frac{(\frac{r}{r_{h}}-1)^{\frac{1}{2}}}{(r-r_{0}%
)^{\frac{r0}{2r_{h}}}}\sinh(\kappa t), \label{72}%
\end{equation}

\begin{equation}
R=\frac{1}{2}(V-U)=\frac{(\frac{r}{r_{h}}-1)^{\frac{1}{2}}}{(r-r_{0}%
)^{\frac{r_{0}}{2r_{h}}}}\cosh(\kappa t), \label{73}%
\end{equation}

From these foregoing equations, we immediately observe that%

\begin{equation}
R^{2}-\Im^{2}=\frac{(\frac{r}{r_{h}}-1)}{(r-r_{0})^{\frac{r_{0}}{r_{h}}}},
\label{74}%
\end{equation}

which means that $R=\pm\Im$ corresponds to the future and past horizons. In
other respects, here $\partial_{\Im}$\ is not a timelike Killing vector for
the metric (71). So, it is profitable to consider the timelike Killing vector
of the metric in the following form%

\begin{equation}
\partial_{\widehat{T}}=N(R\partial_{\Im}+T\partial_{R}), \label{75}%
\end{equation}

where $N$ denotes the normalization constant. It admits a specific value that
the norm of the Killing vector becomes negative unity at $R_{d}$ (19) which is
the outer region of the GMHBH. Therefore, at that specific location the
normalization constant is found to be%

\begin{equation}
N=\left.  \frac{r_{h}-r_{0}}{r_{h}}\sqrt{\frac{ar}{2(r-r_{h})(r-r_{0})}%
}\right\vert _{r=R_{d}}=\frac{a(r_{h}-r_{0})}{r_{h}}, \label{76}%
\end{equation}

Without loss of generality, we may only consider the (1+1) dimensional form of
the KS metric (71) which is%

\begin{equation}
ds^{2}=-\pounds (dT^{2}-dR^{2}), \label{77}%
\end{equation}
The calculation of the HJ method is more straightforward in this case. The HJ
equation (15) for the above metric reads%

\begin{equation}
-\pounds ^{-1}\left[  -\left(  \partial_{\Im}I\right)  ^{2}+\left(
\partial_{R}I\right)  ^{2}\right]  +m^{2}=0, \label{78}%
\end{equation}

This equation implies that the ansatz for the $I$ could be written as%

\begin{equation}
I=\rho(R-\Im)+J(x^{i}), \label{79}%
\end{equation}

For simplicity, we may further set $J(x^{i})=0$ and $m=0.$ Now, the energy can
be defined as%

\begin{equation}
E=-\partial_{\widehat{T}}I, \label{80}%
\end{equation}

which is equivalent to%

\begin{equation}
E=-\frac{a(r_{h}-r_{0})}{r_{h}}(R\partial_{\Im}I+T\partial_{R}I), \label{81}%
\end{equation}

Using the above equation with ansatz (79), one derives the following expression.%

\begin{equation}
\rho(y)=\int\frac{Er_{h}}{a(r_{h}-r_{0})y}dy, \label{82}%
\end{equation}

where $y=R-\Im$. The above expression has a divergence at the horizon $y=0$,
namely $R=\Im$. Thus, it leads to a pole at the horizon which could be
overcome by doing a semi-circular contour of integration in the complex plane.
The result is found to be%

\begin{equation}
{\operatorname{Im}}I=\frac{\pi r_{h}E}{a(r_{h}-r_{0})}. \label{83}%
\end{equation}

which means that the Hawking temperature, $T_{H}=\frac{a(r_{h}-r_{0})}{2\pi
r_{h}}$, is impeccably recovered in the background of the KS metric of the GMHBH.

\section{CONCLUSION}

In this paper, by using the relativistic HJ equation we have studied the HR in
the GMHBH background engendered by the theory of $f(\Re)$ gravity. Today, the
GMHBH has become prominent since it is considered as one of the significant
theoretical astrophysical models in which the dark matter halo and the flat
galactic rotation curves are taken into account. In addition to its naive
coordinates, three different regular coordinate systems which are PG, IEF\ and
KS have been employed throughout the present study. It has been shown in
detail that the computed horizon temperatures via the HJ method exactly
matches with the standard Hawking temperature. Among the ans\"{a}tze that we
have used for the HJ equation (15) in the former sections, the one belonging
to the KS coordinates is different than others. Because in the KS coordinates
the time coordinate is not in a simplex form. To this end, we have first found
a proper timelike Killing vector having a normalization constant $N$ (76) such
that the norm of this Killing vector becomes negative unity at $R_{d}$ (19).
Subsequently, with aid of this Killing vector we have managed to identify an
ansatz $I$ which results in $T_{H}$\ within the process of HJ method. During
this computation, without loss of generality, we have discarded the mass of
the scalar particle and neglected the angular dependence of the HJ equation.

In the PG\ coordinates, we have also considered the back reaction effects in
the PWT\ method for the HR of the GMHBH. The modified tunneling rate (48) has
been computed via the log-area correction to the Bekenstein-Hawking entropy
(49). From this, QG corrected Hawking temperature (i.e., $T_{QG}$) have also
been found.

Finally, it is of interest to extend our analysis to yet another particle
other than spin-$0$, which could be photon and fermion. In other words, it
will be interesting to examine whether Maxwell and Dirac equations
\cite{Chandra} on the GMHBH geometry within the HJ and PWT methods yield the
$T_{H}$ or not. This is going to be our next work in the near future.

\end{document}